\documentclass[sort&compress,5p]{elsarticle}
\usepackage{graphicx}
\usepackage[pdftex,colorlinks,citecolor=blue,bookmarks]{hyperref}
\usepackage{bm}
\begin{document}
\title{Multiple shape coexistence in the nucleus $^{80}$Zr}

\author{Tom\'as R. Rodr\'iguez} 
\address{GSI Helmholtzzentrum
  f\"ur Schwerionenforschung, D-64291 Darmstadt, Germany}
\address{Departamento de F\'isica Te\'orica, Universidad
  Aut\'onoma de Madrid, E-28049 Madrid, Spain} 
\author{J. Luis Egido} 
\address{Departamento de F\'isica Te\'orica, Universidad
  Aut\'onoma de Madrid, E-28049 Madrid, Spain}
\begin{abstract}
We study the low-lying energy spectrum of the \textit{rp}-process waiting point nucleus $^{80}$Zr with state-of-the-art beyond mean field methods with the Gogny D1S interaction. Symmetry restoration and configuration mixing of axial and triaxial shapes are included in the calculations. Five $0^{+}$ states corresponding to different nuclear shapes are obtained below 2.25 MeV and several rotational and $\gamma$- bands built upon them are identified. Nevertheless, these states do not modify the $\beta$-decay half-life having a negligible effect in the \textit{rp}-process. A good agreement with the available experimental data is obtained. 
\end{abstract}
\maketitle
\begin{figure}[tbh]
\begin{center}
  \includegraphics[width=\columnwidth]{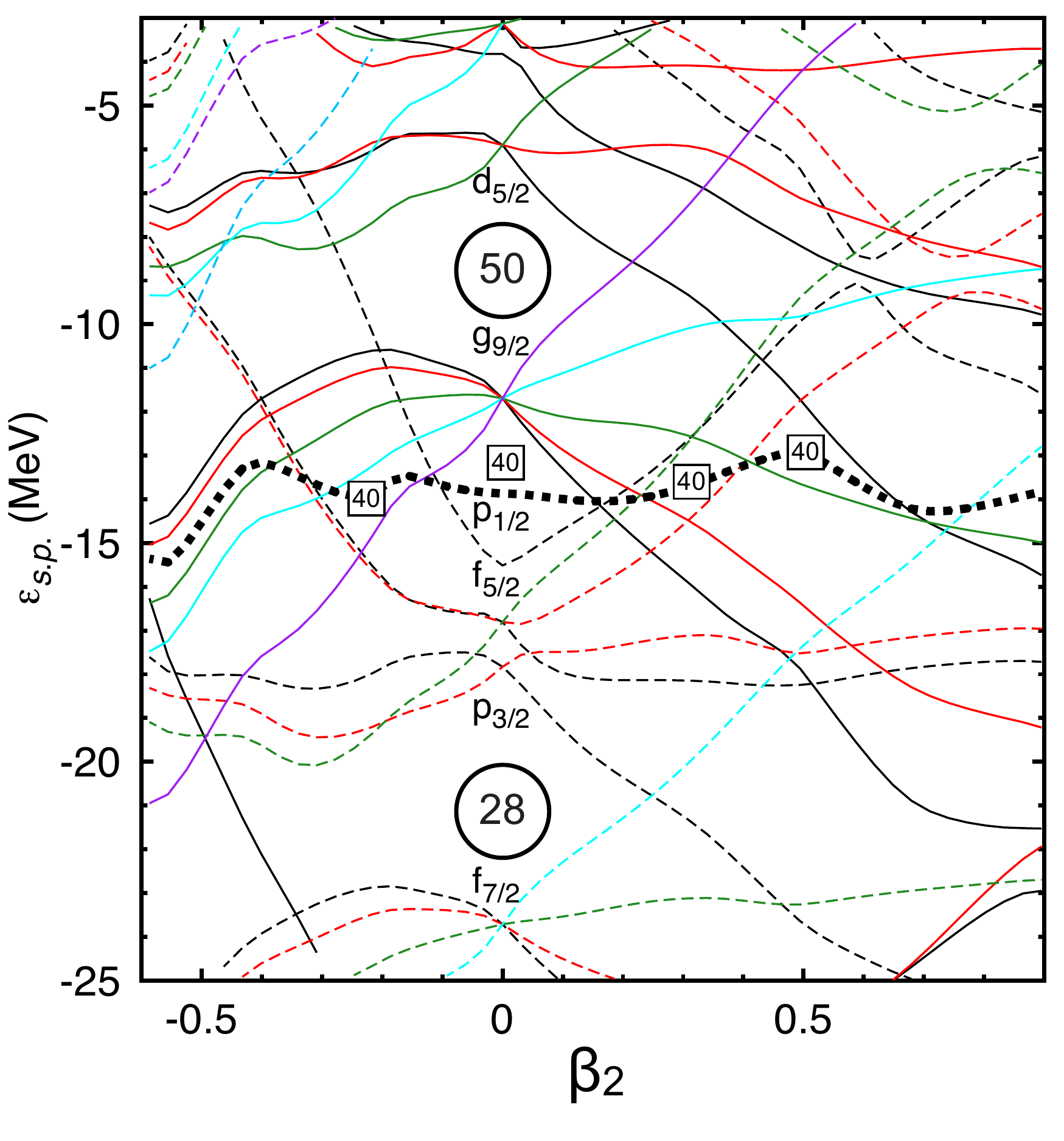}
\end{center}
\caption{(color online) Single particle energies for neutrons (protons follow a similar pattern with energies about 11 MeV higher) as a function of the quadrupole deformation $\beta_{2}$ calculated with the Gogny D1S interaction. The Fermi level is represented by a thick dotted line.}\label{Figure1}
\end{figure}

The study of shape coexistence in nuclear systems is one of the most active topics of current research. In particular, shape coexistence have been observed and/or theoretically predicted in the proton-rich region with mass number $A \approx 80$~\cite{shape-coexistence1,shape-coexistence2,shape-coexistence3,shape-coexistence4,shape-coexistence5,shape-coexistence6,shape-coexistence7,shape-coexistence8}. These nuclei are characterized by having at least one $0^{+}$ excited state in the low lying energy spectrum that can be associated to a shape different from the ground state one. In this region, such an effect is mainly due to the competition between the closure of the $pf$ subshells and the intrusion of some levels of the $g_{9/2}$ and $d_{5/2}$ shells whenever the deformation is increased. Both schematic and self-consistent interactions show this particular behavior~\cite{Moller,Naza,Heyde,skyrme_triax_SIII}. As an example, we display in Fig.~\ref{Figure1} the single particle energies calculated with the Gogny D1S interaction~\cite{Berger84} as a function of the quadrupole deformation $\beta_{2}$.  The Fermi level crosses the $N=Z=40$ spherical gap and, increasing the deformation, the $p_{1/2}$ and $f_{5/2}$ orbits are depopulated while $g_{9/2}$ and $d_{5/2}$ levels enter below the Fermi energy. This structure favors the appearance of different minima in the potential energy surface and the competition and/or coexistence of different shapes at low energies. The nucleus $^{80}$Zr ($N=40$ protons and $Z=40$ neutrons) belongs to this region of potential shape coexistence. \\
Experimentally, there are only six levels measured in this isotope, being identified as members of the ground state rotational band~\cite{Exp_PRL59,Exp_PRL87}. From the theoretical point of view, this isotope has been previously studied using microscopic-macroscopic models~\cite{Moller,Naza,Heyde}, Monte Carlo shell model~\cite{langanke.dean} and self-consistent mean field approximations with Skyrme~\cite{skyrme_triax_SIII,skyrme_Sly4,skyrme_HF_346}, Relativistic~\cite{Relat1,Relat2} and Gogny interactions~\cite{Gogny1}. There are also calculations with these interactions including beyond mean field effects,  like axial angular momentum projection~\cite{Relat3} with additional mixing of axial shapes~  \cite{skyrme_bender}, parity projection\cite{skyrme_tetra} and in the framework of the
collective Hamiltonian~\cite{Gogny2}. Most of these works are focused on the ground state properties of $^{80}$Zr and not on its excitation spectrum.  For the description of shape coexistence several ingredients are required such as the triaxial shapes with the corresponding angular momentum projection and shape mixing~\cite{relat_triax}, and, additionally, the particle number projection~\cite{skyrme_triax}. The latter is important: 1) to remove the possible spurious contributions coming from eigenstates of the particle number operators with different eigenvalues from the desired one, and 2) if the projection is performed before the variation~\cite{AER.PLB,PRL_99}  to  avoid  spurious phase transitions  in the weak pairing regime ( pairing collapse ) and  to provide the optimal energy since one applies the variational principle to a wave function with sharp particle number  at variance with the HFB case. \\
\indent Recently, we have presented ~\cite{rod_triax} a beyond mean field approach, more precisely a Symmetry Conserving Configuration Mixing (SCCM) theory, which considers linear combination of quadrupole deformed wave functions projected to particle  and triaxial angular momentum  quantum numbers. The finite range density dependent Gogny interaction was used and a model application was done to study  the nucleus  $^{24}$Mg with a small configuration space. We now apply this state-of-the-art SCCM theory for the first time to a case requiring a large configuration space, namely the study of the prospective axial and triaxial shape coexistence in the nucleus $^{80}$Zr. The high predictability of the Gogny interaction provides an added bonus to the calculations. As it will be shown below, the approach used here improves the one given in Ref.~\cite{Gogny2} although the underlying interaction is the same. There, particle number projection is not performed and the overlaps needed for the shape mixing are assumed to be gaussian to build the collective hamiltonian. These approaches are valid under certain conditions but not in the whole triaxial plane.\\
\indent Proton-rich nuclei are also of paramount relevance to describe the synthesis of elements in the so-called rapid-proton capture process (\textit{rp}-process) which occurs in X-ray bursts produced in binary systems composed by a massive star and a neutron star~\cite{rp_process,rp_process2}. Along the \textit{rp}-process path there are some even-even $N=Z$  nuclei such as $^{64}$Ge, $^{68}$Se and $^{72}$Kr with a small proton capture cross-section and a relatively long beta decay half-life which are known as waiting points. Hence, final abundances of the elements produced in these bursts and also the time-scale of the nucleosynthesis process depend crucially on the structure of these specific nuclei. The nucleus $^{80}$Zr is one of these  \textit{rp}-process waiting points although the beta decay half-life of its ground state is shorter than in the cases referred above~\cite{Exp_PRL84}. However, the appearance of isomeric states in the low-lying energy spectrum could open additional $\beta^{+}$ decay channels modifying the effective half-life of this nucleus, speeding-up or delaying the \textit{rp}-process.\\
\indent The nuclear shape is a semiclassical concept which can be associated with a mean field description. Therefore the  optimal approach to describe nuclear shapes is a  self-consistent mean field theory (Hartree-Fock-Bogoliubov, HFB, in our case). By constraining the pertinent operators (the quadrupole mass operators $\hat{Q}_{2\mu}$) one is able to produce different shapes at will.  In our case the well known deformation parameters $(\beta_{2},\gamma)$ are defined as:
\begin{eqnarray}
\langle\Phi|\hat{Q}_{20}|\Phi\rangle&=&\frac{\beta_{2}\cos\gamma}{C}\nonumber\\
\langle\Phi|\hat{Q}_{22}|\Phi\rangle&=&\frac{\beta_{2}\sin\gamma}{\sqrt{2}C}\nonumber\\
C&=&\sqrt{\frac{5}{4\pi}}\frac{4\pi}{3r_{0}^{2}A^{5/3}}
\label{betagamma}
\end{eqnarray}
being $r_{0}=1.2$ fm and $A$ the mass number. \\
\indent HFB wave functions (w.f.) -$|\Phi_{\beta_{2}\gamma}\rangle$- are not eigenstates of symmetry operators like the angular momentum (AM) and the particle number (PN) and a restoration of such symmetries is mandatory.
For finite systems the PN symmetry must be imposed from the very beginning and the w.f.'s $|\Phi_{\beta_{2}\gamma}\rangle$ are found by solving the equations resulting of a variation after the PN projection (PN-VAP approximation)~\cite{Anguiano_VAP_01}, i.e., minimizing the projected energy with the mentioned constraints. With these w.f.'s one is able to calculate PN and AM projected potential energy surfaces (PNAMP approach): \begin{equation}
E^{I,N,Z}_{\beta_2,\gamma} = \frac{\langle\Phi_{\beta_2,\gamma}|\hat{H}P^{I,N,Z}|
\Phi_{\beta_2,\gamma}\rangle}{\langle\Phi_{\beta_2,\gamma}|P^{I,N,Z}| \Phi_{\beta_2,\gamma}\rangle}  
\end{equation}
where $I,M,K$ are the AM and the $z$ component of the vector $\vec{I}$ in the laboratory and intrinsic frame respectively and $P^{I,N,Z}=P^{I}_{00}P^NP^Z$ and $P^{I}_{00}$ ($P^{I}_{MK}$ in general) and ${P}^{N(Z)}$ are the AM and the neutron (proton) projectors~\cite{RingSchuck}.\\
The energy eigenstates are not in general eigenstates of the shape operators and a mixing is expected if the local minima are close in energy. Therefore our final approach is a SCCM calculation  (with shape and $K$ mixing) in the framework of the Generator Coordinate Method (GCM):
\begin{equation}
|IM;NZ; \sigma\rangle=\sum_{K\beta_{2}\gamma}f^{I;NZ;\sigma}_{K\beta_{2}\gamma}P^{I}_{MK}P^{N}P^{Z}|\Phi_{\beta_{2}\gamma}\rangle
\label{gcm_state}
\end{equation}
\begin{figure*}[tbh]
\begin{center}
  \includegraphics*[width=\textwidth]{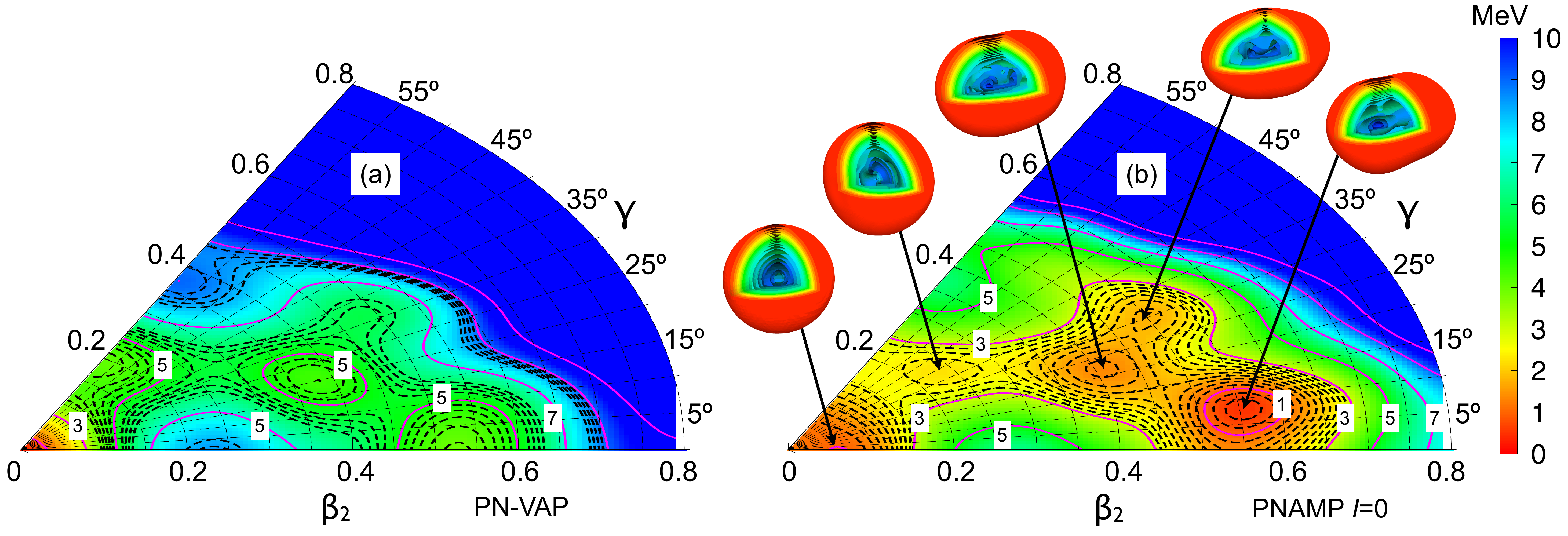}
\end{center}
\caption{(color online) Potential Energy Surfaces -normalized to their absolute minima respectively- calculated for $^{80}$Zr using the Gogny D1S
  interaction: (a) only particle number symmetry is restored -before variation- and (b) the rotational symmetry is also restored ($I=0$), i.e., the PNAMP approach. Spatial densities for each minimum of (b) are also shown. Contour lines are separated by 0.2 MeV (dashed) and 1 MeV (continuous). The color scale is the same for (a) and (b).}\label{Figure2}
\end{figure*}
 $\sigma=1,2,...$ labels the states for a given value of the angular momentum ($\sigma=1$ corresponds to the Yrast states). The coefficients $f^{I;NZ;\sigma}_{K\beta_{2}\gamma}$ in Eq.~\ref{gcm_state} are found by solving the Hill-Wheeler-Griffin equations that are deduced from the minimization of the energy calculated with the non-orthogonal set of wave functions $\lbrace P^{I}_{MK}P^{N}P^{Z}|\Phi_{\beta_{2}\gamma}\rangle\rbrace$~\cite{RingSchuck}. The solution of these equations provides the energy levels and all the information needed for calculating, for example, reduced transition probabilities, the so-called collective wave functions (square of the probability density, normalized to 1, of finding the state $(I, \sigma)$ with given deformation parameters $(\beta_{2},\gamma)$), decomposition of the states in $K$, etc. More details about how the observables are computed within this framework can be found in Reference~\cite{rod_triax} (and references therein). In this work, HFB  states $|\Phi_{\beta_{2}\gamma}\rangle$  conserve parity, time-reversal and simplex symmetries and they are expanded in a spherical harmonic oscillator basis which includes nine major oscillator shells. Concerning the input parameters chosen in these calculations, we use $N_{\mathrm{GCM}}=60$ intrinsic states to form a triangular mesh in the $(\beta_{2},\gamma)$ plane. Finally, although the exploration of the octupole degree of freedom is beyond the scope of the present letter, we have checked the influence of $\beta_{3}$  performing axial HFB calculations in the range of quadrupole deformations studied here. As a result, we do not find any solution where the parity symmetry is broken.  \\
\indent To study the possible shape coexistence in this nucleus we start with analyzing the potential energy surface (PES) in the triaxial $(\beta_{2},\gamma)$ plane calculated with the PN-VAP approximation. In Fig.~\ref{Figure2}(a) we observe a very rich structure with five minima located at deformations $(\beta_{2},\gamma)=$ $(0,0^{\circ})$, $(0.53,0^{\circ})$, $(0.37,20^{\circ})$, $(0.47,32^{\circ})$ and $(0.2,60^{\circ})$. The axial minima ($\gamma=0^{\circ},60^{\circ}$) can be directly related to the gaps observed in the single-particle energies represented in Fig~\ref{Figure1}. In this approach, the absolute minimum corresponds to the spherical point which is in contradiction to the rotational behavior observed experimentally. The other minima are, respectively,  at 4.0, 4.4, 5.7 and 4.1 MeV above the spherical one. Furthermore, this energy landscape displays a rather gamma-soft behavior. The barriers between the prolate, the triaxial and the lowest oblate minimum are smaller than 1 MeV. Nevertheless, this structure is expected to change whenever beyond mean field effects, including angular momentum projection and shape mixing, are taken into account. \\
In Fig.~\ref{Figure2}(b) the PES in the PNAMP approach ($I=0$) is shown. We also represent the spatial density distribution of the intrinsic states that correspond to the different minima found in the surface to give a better idea about their shapes. We observe that, except for the spherical point, the rotational correction induced by the symmetry restoration produces a significant lowering of the energy in the whole plane. The prolate and oblate minima in the PN-VAP approach have been displaced to triaxial shapes and slightly larger $\beta_{2}$ deformations. In fact, the absolute minimum corresponds now to a moderately triaxial deformed shape $(\beta_{2},\gamma)=(0.55,7^{\circ})$ while the spherical one has been displaced to a slightly prolate shape, being now at $\approx 1.4$ MeV above the absolute minimum. This result goes in the right direction to reproduce the experimental data. The effect is somewhat similar to the $^{32}_{12}$Mg$_{20}$ case in the island of inversion where the deformed configuration becomes the ground state only after restoring the rotational symmetry~\cite{Rayner}. Nevertheless, we have to include the possible influence of shape and $K$ mixing in the calculation because the barriers between neighboring triaxial minima remain of the order of 1 MeV or less.
\begin{figure*}[htb]
\begin{center}
  \includegraphics*[width=\textwidth]{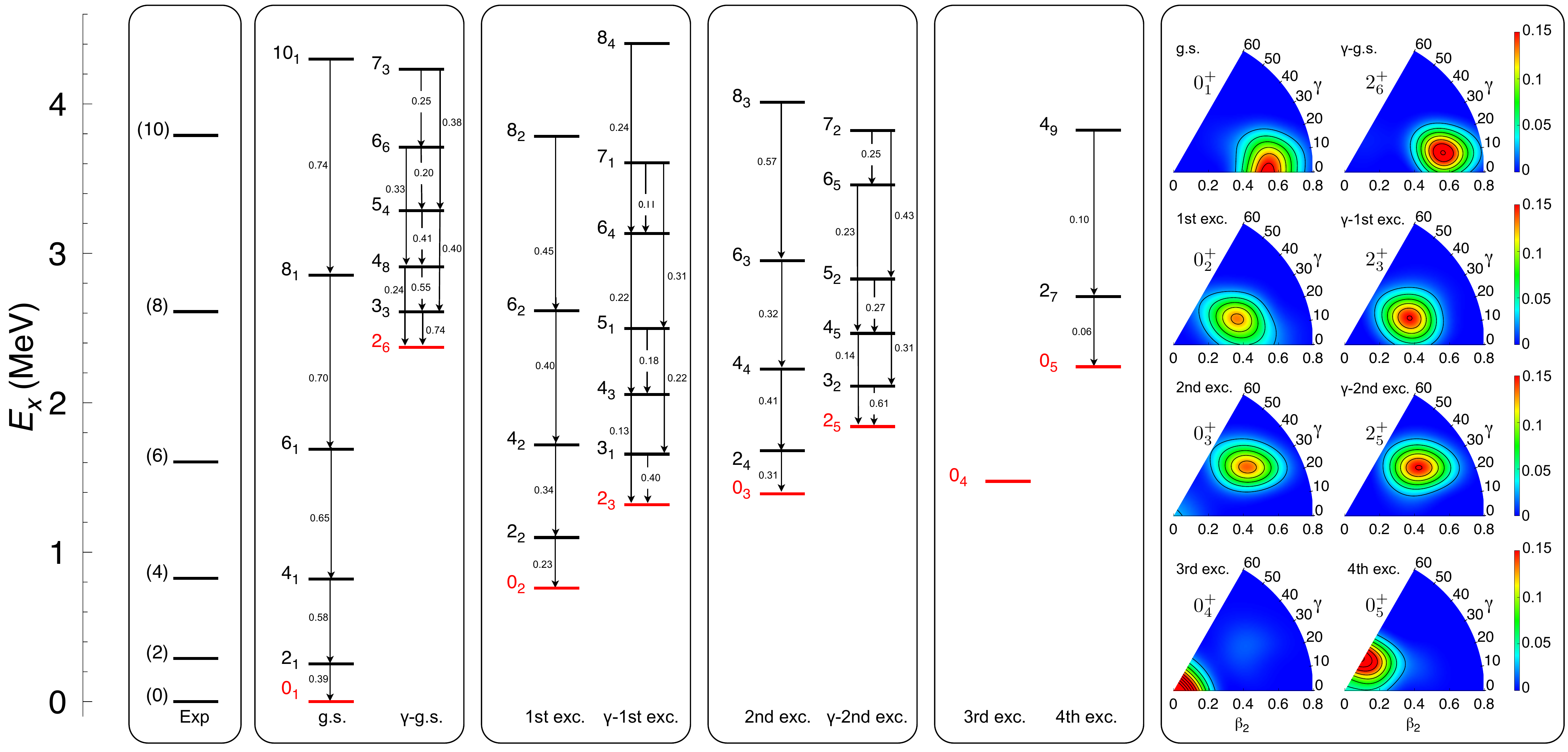}
\end{center}
\caption{(color online) Left: Experimental \cite{Exp_PRL59,Exp_PRL87} and theoretical spectra for $^{80}$Zr. $B(E2)$ values are given in $e^{2}b^{2}$ Right: Square of the collective wave functions for the band heads.}
  \label{Figure3}
\end{figure*}
Hence, we perform SCCM calculations including particle number and angular momentum projection (Eq. \ref{gcm_state}) to obtain the energy levels and electric quadrupole reduced transition probabilities $B(E2)$. The predicted spectrum is shown in Fig.~\ref{Figure3}. The states are sorted in bands according to the values of the highest transition probabilities $B(E2)$ between those levels. Highest values of $B(E2)$ within the different bands are also displayed in Fig.~\ref{Figure3} while several interband transitions are given in Table~\ref{table1}. Additionally, the distribution of probability of having a given deformation $(\beta_{2},\gamma)$ for the band-head states is also plotted on the right in Fig.~\ref{Figure3}. This distribution corresponds to the square of the collective wave functions in the triaxial plane. Here we observe very well defined maxima showing that the concept of shape is still valid even after the shape mixing.  The energy ordering of these maxima coincides with the minima of Fig.~\ref{Figure2}(b) with the exception of the spherical and the oblate ones which have been exchanged. Interestingly, the shape fluctuations bring back the "almost" prolate and oblate minima to pure prolate and oblate  shapes as in Fig.~\ref{Figure2}(a).
In addition, the distribution of probability remains practically the same for the rest of the members of a particular band and corroborates the stability of the different minima. The distribution in $K$ for the states belonging to a band is also similar as it is shown in Table~\ref{table2}. Furthermore, the most intense electric quadrupole transitions, both between intraband and interband states, are related to similar distribution in $K$ of the initial and final states (see Tables~\ref{table1},~\ref{table2} and Fig.~\ref{Figure3}).\\
\indent In Fig.~\ref{Figure3}, on the left, we observe four $0^{+}$ excited states below 2.25 MeV, built on top of different shapes, coexisting with the ground state.
The ground state band is predicted to be a rotational band built on top of an axial prolate deformed configuration with the usual sequence $I=0,2,4,6,...$ (g.s. in Figs.~\ref{Figure3} and \ref{Figure2}(b)). The collective wave functions have their maxima at $(\beta_{2},\gamma)=(0.55,0^{\circ})$ that corresponds to the minimum found in Fig.~\ref{Figure2}. Comparing these results with the experimental energies we obtain a rather good agreement. This shows again the relevance of restoring the rotational invariance in this nucleus where a spherical ground state is found at the PN-VAP approach.  The theoretical values of the energies tend to follow a more rotational behavior $I(I+1)$ than the data and the value for the $B(E2, 2_{1}\rightarrow0_{1})$ seems to be overestimated according to the trend of the experimental data along the isotonic $N=40$ chain~\cite{DataBase}. These effects could be related to the fact that we are not including time-reversal symmetry breaking intrinsic states. The angular momentum projection of cranked states would compress the spectrum -affecting only the excitation energies for states with $I\neq0$- and would modify also the value of the transition probabilities~\cite{crank_skyrme}.
All the states belonging to the g.s. band correspond to rather pure $K=0$ configurations (see Table~\ref{table2}). According to a semiclassical picture~\cite{RingSchuck}, a $\gamma$-vibrational band with $K=2$ is expected to be present with an $I=2^{+}$ state as the band head and $\Delta I=1$. These states are found indeed at $~2.4$ MeV as we can see in the figure ($\gamma$-g.s. in Fig.~\ref{Figure3}, see also Table~\ref{table2}). The transition probabilities along this band are large and, since $K$-mixing is very weak, those to the g.s. band very small. \\
The first excited band built on top of the $0^{+}_{2}$ state ($E_{0^{+}_{2}}=0.76$~MeV) has also a rotational character based on a triaxial configuration (1st exc. in Figs.~\ref{Figure3} and ~\ref{Figure2}(b)). The distribution of probability is concentrated at $(\beta_{2},\gamma)\approx(0.4,20^{\circ})$. As we see in Table~\ref{table2}, this is mostly a $K=0$ band although some mixing with $K=2$ components is present, especially in the $I=6^{+}_{2},8^{+}_{2}$ states. We can also find a quasi-$\gamma$ band partner with the  $2^{+}_{3}$ state  at 1.32 MeV as the band head, and  with the probability peaked at the same deformation as its partner ($\gamma$-1st exc. and Fig.~\ref{Figure3}). The even $I$ states of this band have a considerable $K=0,2$ mixing while the odd $I$ levels are rather pure $K=2$ states. Correspondingly, we obtain large  $B(E2)$ values for $\Delta I=2$ transitions between the levels of this band and rather strong inter-band transitions to the first excited band. \\
We also find another band and its quasi-$\gamma$ band partner with strong $K$ mixing built on top of the triaxial minimum of the PES at $(\beta_{2},\gamma)=(0.5,30^{\circ})$ (2nd exc. and $\gamma$-2nd exc. in Figs.~\ref{Figure3} and \ref{Figure2}(b)). The excitation energies of the band heads are $E(0^{+}_{3})=1.4$ and $E(2^{+}_{5})=1.8$ MeV respectively. In the quasi-$\gamma$ band the largest values for the $B(E2)$ correspond to $\Delta I=2$ and also $\Delta I=1$  transitions with an odd $I$  initial state. The states of these two bands are also strongly connected to the first excited quasi-$\gamma$ band. Finally, we obtain two more $0^{+}$ states located on top of the spherical and oblate minima of the PES at excitation energies of 1.5 and 2.24 MeV, respectively. \\
\renewcommand {\arraystretch}{1.5}
\begin{table}[t]
\centering
\scriptsize
\begin{tabular}{cc||cc} 
\hline \hline
$I_{\sigma_i} \rightarrow I_{\sigma_f} $ & $B(E2)$ (e$^{2}$b$^{2}$) & $I_{\sigma_i} \rightarrow I_{\sigma_f}$ & $B(E2)$ (e$^{2}$b$^{2}$)\\
\hline
$2_{6}^{gs\gamma}\rightarrow2_{1}^{gsb}$ & 0.017 & $0_{2}^{1eb}\rightarrow2_{1}^{gsb}$ & 0.039 \\
$2_{3}^{1e\gamma}\rightarrow2_{2}^{1eb}$ & 0.247 & $0_{3}^{2eb}\rightarrow2_{3}^{1e\gamma}$ & 0.055 \\
$6_{3}^{2eb}\rightarrow4_{3}^{1e\gamma}$ & 0.220 &  $0_{3}^{2eb}\rightarrow2_{2}^{1e\gamma}$ & 0.018 \\
$4_{8}^{gs\gamma}\rightarrow4_{1}^{gsb}$ & 0.021&  $0_{4}^{3eb}\rightarrow2_{2}^{1eb}$ & 0.010  \\
$2_{5}^{2e\gamma}\rightarrow2_{4}^{2eb}$ & 0.441  & $0_{5}^{4eb}\rightarrow2_{2}^{1eb}$ & 0.023\\
\hline
\hline  
\end{tabular}
\caption{Selected theoretical interband electric quadrupole reduced transition probabilities $B(E2)$. Following Fig.~\ref{Figure3} the short notation {\it gsb} stands
for ground state band, {\it 1eb} for 1st excited band,  $1e\gamma$ for 1st excited $\gamma$-band, etc.}
\label{table1}
\end{table}
\renewcommand {\arraystretch}{1}
\begin{table}[t]
   \centering
\scriptsize
   \begin{tabular}{cccc||cccc} 
   \hline \hline
$I_{\sigma}$ & $K=0$ & $|K|=2$ & $|K|=4$ &$I_{\sigma}$ & $K=0$ & $|K|=2$ & $|K|=4$ \\
\hline
$0_{1}$ & 1.000 & -- & -- & $2_{6}$ & 0.022 & 0.978 & --  \\
$2_{1}$ & 0.997 & 0.003	& -- & $3_{3}$ & 0.000 & 1.000 & -- \\
$4_{1}$ & 0.995 & 0.004 & 0.001 & $4_{8}$ & 0.005 & 0.980 & 0.015  \\
$6_{1}$ & 0.990 & 0.010 & 0.000 & $5_{4}$ & 0.000 & 0.976 & 0.024  \\
\hline
$0_{2}$ & 1.000 & -- & -- & $2_{3}$ & 0.210 & 0.790 & -- \\
$2_{2}$ & 0.889 & 0.111 & -- & $3_{1}$ & 0.000 & 1.000 & -- \\
$4_{2}$ & 0.804 & 0.191 & 0.005 & $4_{3}$ & 0.352 & 0.613 & 0.035  \\
$6_{2}$ & 0.725 & 0.260 & 0.014 & $5_{1}$ & 0.000 & 0.946 & 0.054  \\
$8_{2}$ & 0.638 & 0.326 & 0.034 & $6_{4}$ & 0.201 & 0.767 & 0.024 \\
\hline
$0_{3}$ & 1.000 & -- & -- & $2_{5}$ & 0.267 & 0.733 & -- \\
$2_{4}$ & 0.707 & 0.293 & -- & $3_{2}$ & 0.000 & 1.000 & -- \\
$4_{4}$ & 0.462 & 0.498 & 0.040 & $4_{5}$ & 0.239 & 0.312 & 0.449 \\ 
$6_{3}$ & 0.570 & 0.369 & 0.058 & $5_{2}$ & 0.000 & 0.770 & 0.230 \\
$8_{3}$ & 0.525 & 0.383 & 0.084 & $6_{5}$ & 0.262 & 0.069 & 0.581 \\ 
\hline
$0_{5}$ & 1.000 & -- & -- & & & &\\
$2_{7}$ & 0.777 & 0.223 & -- & & & &\\
$4_{9}$ & 0.808 & 0.095 & 0.097 & & & &\\
\hline
\hline
   \end{tabular}
   \caption{Decomposition of the collective states appearing in Figure~\ref{Figure3} in their $|K|\leq4$ components}
   \label{table2}
\end{table}
\normalsize
From the astrophysical point of view, the appearance of isomeric states in the low-lying energy spectrum could modify the effective $\beta^{+}$ decay half-life towards the $^{80}$Y isotope. However, we obtain half-lives of the different excited $0^{+}$  states much shorter than the typical half-lives for a $\beta^{+}$ decay process. Furthermore, the excitation energies of these states are too high to allow for their population at the typical temperatures of the $rp$-process $T\sim1$ GK (equivalent to $E\sim100$ keV)~\cite{rp_process2}. Therefore, the $\beta^{+}$ decay of $^{80}$Zr is expected to occur from its ground state only which has an experimental half-life of $\sim4.1$~s~\cite{Exp_PRL84}. 

In summary, we have applied state-of-the-art beyond mean field methods to study the structure of the \textit{rp}-process waiting point nucleus $^{80}$Zr. We find five $0^{+}$ states below 2.25 MeV that correspond to different minima in the triaxial potential energy landscape which constitutes a unique example of multiple shape coexistence in the nuclear chart. This feature only occurs if full triaxial angular momentum projection is performed.   However, we found that due to their   short lifetimes the effect of the excited $0^{+}$ states on the beta-decay of $^{80}$Zr and therefore on the \textit{rp}-process is negligible. An experimental observation of the excited $0^{+}$ states predicted by our calculations,
once radioactive beam intensities will be high enough to perform for example Coulomb excitation
studies of $^{80}$Zr, will certainly be of highest interest.

The authors acknowledge financial support from the Spanish Ministerio de Educaci\'on y Ciencia 
under contract  FPA2009-13377-C02-01. T.R.R. thanks G. Mart\'inez-Pinedo and C. Domingo-Pardo for helpful discussions and fundings from the Programa de Ayudas para Estancias de Movilidad Posdoctoral 2008 and a HIC4FAIR scholarship.

\end{document}